\documentclass[preprintnumbers,prd,showpacs,floatfix,superscriptaddress,nofootinbib,twocolumn, letterpaper]{revtex4-1}
\pdfoutput=1

\usepackage{longtable}
\usepackage{bm}
\usepackage{relsize}
\usepackage{amsfonts}
\usepackage{amsmath}
\usepackage{amssymb,epsf}
\usepackage{latexsym}
\usepackage{graphicx,epsfig}
\usepackage{amssymb}
\usepackage{float}
\usepackage{subfigure}
\usepackage{epstopdf}
\usepackage[colorlinks=true,citecolor=blue,linkcolor=blue,urlcolor=black]{hyperref}
\usepackage{dcolumn}
\usepackage{psfrag}
\usepackage{wrapfig}
\usepackage{makeidx}
\usepackage{epsf}
\usepackage{color}
\usepackage{multirow}
\usepackage{mathtools}

\begin{document}

\title{Non-exotic traversable wormholes in $f\left(R,T_{ab}T^{ab}\right)$ gravity}

\author{Jo\~{a}o Lu\'{i}s Rosa}
\email{joaoluis92@gmail.com}
\affiliation{University of Gda\'{n}sk, Jana Ba\.{z}y\'{n}skiego 8, 80-309 Gda\'{n}sk, Poland}
\affiliation{Institute of Physics, University of Tartu, W. Ostwaldi 1, 50411 Tartu, Estonia}

\author{Nailya Ganiyeva}
\email{nailhask8@hotmail.com}
\affiliation{Instituto de Astrof\'{i}sica e Ci\^{e}ncias do Espaço, Faculdade de Ci\^{e}ncias da Universidade de Lisboa,
Edif\'{i}cio C8, Campo Grande, P-1749-016 Lisbon, Portugal}

\author{Francisco S. N. Lobo}
\email{fslobo@fc.ul.pt}
\affiliation{Instituto de Astrof\'{i}sica e Ci\^{e}ncias do Espaço, Faculdade de Ci\^{e}ncias da Universidade de Lisboa,
Edif\'{i}cio C8, Campo Grande, P-1749-016 Lisbon, Portugal}
\affiliation{Departamento de F\'{i}sica, Faculdade de Ci\^{e}ncias da Universidade de Lisboa, Edif\'{i}cio C8, Campo Grande, P-1749-016 Lisbon, Portugal}

\date{\today}

\begin{abstract} 
In this work we analyze traversable wormhole spacetimes in the framework of a covariant generalization of Einstein's General Relativity known as energy-momentum squared gravity, or $f\left(R,\mathcal T\right)$ gravity, where $R$ is the Ricci scalar, $\mathcal T=T_{ab}T^{ab}$, and $T_{ab}$ is the energy-momentum tensor. Considering a linear $f\left(R,\mathcal T\right)=R+\gamma \mathcal T$ form, we show that a wide variety of wormhole solutions for which the matter fields satisfy all the energy conditions, namely the null, weak, strong and dominant energy conditions, exist in this framework, without the necessity for a fine-tuning of the free parameters that describe the model. Due to the complexity of the field equations these solutions are obtained through an analytical recursive algorithm. A drawback of the solutions obtained is that they are not naturally localized, and thus a matching with an external vacuum is required. For that purpose, we derive the junction conditions for the theory, and we prove that a matching between two spacetimes must always be smooth, i.e., no thin-shells are allowed at the boundary. Finally, we use these junction conditions to match the interior wormhole spacetime to an exterior vacuum described by the Schwarzschild solution, thus obtaining traversable localized static and spherically symmetric wormhole solutions satisfying all energy conditions for the whole spacetime range. We also prove that the methods outlined in this work can be straightforwardly generalized to more complicated dependencies of the action in $\mathcal T$, as long as crossed terms between $R$ and $\mathcal T$ are absent.
\end{abstract}

\maketitle

\section{Introduction}\label{sec:intro}

A wormhole is a topological object connecting two spacetime manifolds or two regions of the same spacetime manifold. These objects have been widely studied in the framework of General Relativity (GR) \cite{morris1,Morris:1988tu,visser1,lemos1,Visser:2003yf,Kar:1995ss}, but they feature an important drawback: the requirement that the wormhole spacetime be traversable entails the flaring-out condition   \cite{morris1}, which through the Einstein field equations violates the null energy condition (NEC), and consequently violates all of the energy conditions \cite{visser1,Hawking:1973uf}. A matter distribution that violates the NEC is thus denoted as \textit{exotic}, and is of limited physical relevance due to the rarity of an experimental counterpart.

To overcome the limitation of GR stated above, one usually recurs to the analysis of wormhole 
spacetimes in the framework of modified theories of gravity 
\cite{agnese1,nandi1, camera1,camera2,lobo1,garattini1,lobo2,garattini2,lobo3,MontelongoGarcia:2011ag,garattini3,myrzakulov1,lobo4}, where the additional components of the gravitational sector preserve the geometry of the wormhole throat traversable, while keeping the matter components non-exotic. Such a result can be attained in several distinct modifications of GR, from $f\left(R\right)$ gravity and its extensions \cite{lobo5,capozziello1,rosa1,rosa2,rosalol,rosalol2,kull1}, to couplings between curvature and matter \cite{garcia1,garcia2}, theories with additional fundamental fields \cite{harko1,anchordoqui1}, Gauss-Bonnet gravity \cite{bhawal1,dotti1,mehdizadeh1}, and braneworld scenarios \cite{bronnikov1,lobo6}.

In this work, we are interested in a covariant generalization of Einstein's GR known as energy-momentum squared gravity, or $f\left(R,\mathcal T\right)$ gravity \cite{katirci1,roshan1}, where $\mathcal T=T_{ab}T^{ab}$.  This particular modification of GR possesses similar features to the previously explored curvature-matter coupling theories \cite{Bertolami:2007gv,Harko:2010mv,Harko:2011kv,Haghani:2013oma}, where the energy-momentum tensor is not conserved.
$f\left(R,\mathcal T\right)$ gravity has been studied in a wide variety of topics from cosmological models \cite{board1,akarsu1,bahamonde1,akarsu3,akarsu4,barbar1} to compact objects \cite{nari1,akarsu2,singh1,sharif1}, including black-holes \cite{chen1,rudra1}. However, the literature regarding wormhole physics in this theory is scarce: a few specific solutions were found through a Noether symmetry approach, but these solutions lack physical relevance in the sense that they violate the NEC and, consequently, all the other more restrictive energy conditions \cite{sharif2,sharif3}. The main objective of this work is thus to suppress this literary gap and provide a clear analysis of physically relevant traversable wormhole spacetimes in $f\left(R,\mathcal T\right)$ gravity.

In the pursuit of physically relevant spacetime solutions describing localized objects, one frequently makes use of the so-called junction conditions. The junction conditions for GR were derived long ago \cite{israel1} and proven useful in several astrophysical contexts, e.g. the analysis of traversable wormholes \cite{visser2,visser3,Lobo:2004rp,Lobo:2004id,Lobo:2005us,Lobo:2005yv}, fluid stars \cite{schwarzschild1,rosafluid,rosafluid2} and gravitational collapse \cite{oppenheimer1,rosa112}. The junction conditions are theory-dependent, and several works analyze these conditions in modified theories of gravity, from $f\left(R\right)$ gravity and its extensions \cite{senovilla1,Vignolo:2018eco,Reina:2015gxa,Deruelle:2007pt,Olmo:2020fri,rosafrt,rosafrt2} to theories with additional fundamental fields \cite{suffern,Barrabes:1997kk,Padilla:2012ze} and metric-affine gravity \cite{delaCruz-Dombriz:2014zaa,Arkuszewski:1975fz,amacias}. A second objective of this work is thus to provide the junction conditions of linear $f\left(R,\mathcal T \right)$ along with explicit examples of application.

This paper is organized as follows. In Sec. \ref{sec:theory}, we introduce the $f\left(R,\mathcal T\right)$ theory of gravity along with our assumptions for the gravitational and matter sectors. In Sec. \ref{sec:worms}, we solve the field equations and obtain solutions for non-exotic traversable wormhole spacetimes. In Sec. \ref{sec:matching}, we derive the junction conditions for linear $f\left(R,\mathcal T\right)$ gravity and perform a matching between the interior wormhole region previously obtained and an exterior vacuum spacetime. In Sec. \ref{sec:extensions}, we analyze extensions of the theory to higher-order powers of $\mathcal T$. Finally, in Sec. \ref{sec:concl} we present our conclusions.

\section{Theory and framework}\label{sec:theory}

\subsection{Action and field equations of $f\left(R,\mathcal T\right)$}

In this work, we are interested in studying wormholes in the $f\left(R,\mathcal T\right)$ theory of gravity. The action $S$ that describes such a theory can be written as
\begin{equation}\label{geo_action}
S=\frac{1}{2\kappa^2}\int_\Omega \sqrt{-g}f\left(R,\mathcal T\right)d^4x+\int_\Omega \sqrt{-g}\mathcal L_m d^4x,
\end{equation}
where $\kappa^2=8\pi G/c^4$, with $c$ the speed of light and $G$ the gravitational constant, $\Omega$ is a spacetime manifold described by a set of coordinates $x^a$, $g$ is the determinant of the metric $g_{ab}$, $f\left(R,\mathcal T\right)$ is an arbitrary function of both the Ricci scalar $R=g^{ab}R_{ab}$, with $R_{ab}$ the corresponding Ricci tensor, and the scalar $\mathcal T=T_{ab}T^{ab}$, with $T_{ab}$ the energy-momentum tensor, and $\mathcal L_m$ is the matter Lagrangian. The energy-momentum tensor is defined in terms of the variation of the matter Lagrangian $\mathcal L_m$ with respect to the metric as
\begin{equation}\label{def_tab}
T_{ab}=-\frac{2}{\sqrt{-g}}\frac{\delta\left(\sqrt{-g}\mathcal L_m\right)}{\delta g^{ab}}.
\end{equation}
Throughout this work, we assume a geometrized unit system in which $c=G=1$, and hence $\kappa^2=8\pi$.

Applying the variational method to Eq. \eqref{geo_action} with respect to the metric $g_{ab}$ leads to the modified field equations as
\begin{equation}\label{geo_field}
f_R R_{ab}-\frac{1}{2}g_{ab}f-\left(\nabla_a\nabla_b-g_{ab}\Box\right)f_R=8\pi T_{ab}-f_\mathcal T \Theta_{ab},
\end{equation}
where we have introduced the notation $f_R\equiv \partial f/\partial R$ and $f_\mathcal T=\partial f/\partial \mathcal T$, $\nabla_a$ represents the covariant derivatives and $\Box=g^{ab}\nabla_a\nabla_b$ represents the d'Alembert operator, both in terms of the metric $g_{ab}$, and $\Theta_{ab}$ is an auxiliary quantity arising from the variation of $T_{ab}$ as
\begin{equation}
\Theta_{ab}=\frac{\delta\mathcal T}{\delta g^{ab}}.
\end{equation}
Upon specifying a choice for the matter Lagrangian $\mathcal L_m$ or, equivalently, a choice for the energy-momentum tensor $T_{ab}$, the explicit form of the auxiliary tensor $\Theta_{ab}$ is set. Finally, taking the covariant derivative of Eq. \eqref{geo_field}, one obtains the conservation equation
\begin{equation}\label{geo_conservation}
8\pi \nabla_bT^{ab}=\nabla_b\left(f_\mathcal T \Theta^{ab}\right)+f_R\nabla_bR^{ab}-\frac{1}{2}g^{ab}\nabla_bf.
\end{equation}
This result implies that the energy-momentum tensor $T_{ab}$ is no longer required to be conserved in this theory, i.e., in general one has $\nabla_b T^{ab}\neq0$, a notable difference in comparison with GR.

For the purpose of this work, we assume that the function $f\left(R,\mathcal T\right)$ is separable and linear in both $R$ and $\mathcal T$, i.e., we write 
\begin{equation}
    f\left(R,\mathcal T\right)=R+\gamma \mathcal T,
\end{equation}
where $\gamma$ is a coupling constant. Under this assumption, the field equations in Eq. \eqref{geo_field} and the conservation equation in Eq. \eqref{geo_conservation} reduce to
\begin{equation}\label{field}
G_{ab}=8\pi T_{ab}-\gamma\left(\Theta_{ab}-\frac{1}{2}g_{ab}\mathcal T\right),
\end{equation}
\begin{equation}\label{conservation}
8\pi \nabla_b T^{ab}=\gamma\nabla_b\left(\Theta^{ab}-\frac{1}{2}g^{ab}\mathcal T\right),
\end{equation}
respectively, where we have introduced the Einstein tensor $G_{ab}=R_{ab}-\frac{1}{2}R g_{ab}$. Extensions of these assumptions are addressed below in Sec.  \ref{sec:extensions}.

\subsection{Wormhole metric and matter distribution}

In this work we are interested in static and spherically symmetric traversable wormhole solutions. A general static and spherically symmetric metric can be written in the usual spherical coordinates $\left(t,r,\theta,\varphi\right)$ as
\begin{equation}\label{def_metric}
ds^2=-e^{\zeta\left(r\right)}dt^2+\left[1-\frac{b\left(r\right)}{r}\right]^{-1}dr^2+r^2d\Omega^2,
\end{equation}
where $\zeta\left(r\right)$ is the redshift function, $b\left(r\right)$ is the shape function, and $d\Omega^2=d\theta^2+\sin^2\theta d\varphi^2$ is the solid angle surface element. For the wormhole to be traversable, the functions $\zeta\left(r\right)$ and $b\left(r\right)$ must satisfy a few conditions. First, it is necessary that the spacetime is free of event horizons, to allow an observer to cross the interior of the wormhole without being trapped inside. For this purpose, the redshift function must remain finite throughout the whole spacetime, i.e., $|\zeta\left(r\right)|<\infty$. Furthermore, one needs to impose a geometrical condition at the wormhole throat $r=r_0$, known as the flaring-out condition\footnote{The flaring-out condition in the neighbourhood of the throat takes the form $(b- b'r)/b^2 > 0$ \cite{morris1}}, which can be described by the two following boundary conditions on the shape function:
\begin{equation}\label{def_flaring}
b\left(r_0\right)=r_0, \qquad b'\left(r_0\right)<1.
\end{equation}
Two broad families of solutions for the functions $\zeta\left(r\right)$ and $b\left(r\right)$ that satisfy the requirements above are
\begin{equation}\label{zbfunctions}
\zeta\left(r\right)=\zeta_0\left(\frac{r_0}{r}\right)^\alpha, \qquad b\left(r\right)=r_0\left(\frac{r_0}{r}\right)^\beta,
\end{equation}
where $\zeta_0$ is an arbitrary constant to be specified in what follows, and $\alpha$ and $\beta$ are arbitrary positive exponents. 

Regarding the matter sector, we assume that the distribution of matter is well described by an anisotropic perfect fluid, i.e., the energy-momentum tensor $T_{ab}$ takes the form
\begin{equation}\label{def_matter}
T_a^b=\text{diag}\left(-\rho,p_r,p_t,p_t\right),
\end{equation}
where $\rho\equiv\rho\left(r\right)$ is the energy density, $p_r\equiv p_r\left(r\right)$ is the radial pressure, and $p_t\equiv p_t\left(r\right)$ is the tangential pressure. These quantities are assumed to depend solely on the radial coordinate $r$ to preserve the spherical symmetry of the wormhole. Under these assumptions, the matter Lagrangian takes the form $\mathcal L_m=\frac{1}{3}\left(p_r+2p_t\right)$, and consequently the auxiliary tensor $\Theta_{ab}$ reads
\begin{equation}\label{def_theta_2}
\Theta_{ab}=-\frac{2}{3}\left(p_r+2p_t\right)\left(T_{ab}-\frac{1}{2}g_{ab} T\right)-T T_{ab}+2T_{a}^cT_{cb},
\end{equation}
where $T=g^{ab}T_{ab}$ is the trace of the energy-momentum tensor.

\section{Wormhole solutions}\label{sec:worms}

Under the assumptions outlined above, the field equations in Eq. \eqref{field} feature three independent components, which take the forms:
\begin{eqnarray}\label{eqrho}
8\pi\rho &=& \frac{\gamma}{6}\left(p_r^2-2p_t^2-3\rho^2-8p_rp_t-8p_r\rho-16p_t\rho\right)
	\nonumber \\
&&-\frac{\beta}{r^2}\left(\frac{r_0}{r}\right)^{\beta+1},
\end{eqnarray} 
\begin{eqnarray}\label{eqpr}
8\pi p_r &=& \frac{\gamma}{6}\left(p_r^2+2p_t^2-3\rho^2-12p_rp_t+4p_r\rho-4p_t\rho\right)
	\nonumber \\
&& \hspace{-0.5cm} -\frac{1}{r^2}\left(\frac{r_0}{r}\right)^{\beta+1}-\frac{\alpha\zeta_0}{r^2}\left(\frac{r_0}{r}\right)^\alpha\left[1-\left(\frac{r_0}{r}\right)^{\beta+1}\right],
\end{eqnarray}
\begin{eqnarray}\label{eqpt}
8\pi p_t &=&-\frac{\gamma}{6}\left(p_r^2+6p_t^2+3\rho^2+2p_rp_t+2p_r\rho-2p_t\rho\right)
	\nonumber \\
&& +\frac{1+\beta}{2r^2}\left(\frac{r_0}{r}\right)^{\beta+1}
+\frac{\alpha^2\zeta_0^2}{4r^2}\left(\frac{r_0}{r}\right)^{2\alpha}\left[1-\left(\frac{r_0}{r}\right)^{\beta+1}\right]
\nonumber \\
&&	+\frac{\alpha\zeta_0}{4r^2}\left(\frac{r_0}{r}\right)^\alpha\left[2\alpha-\left(1+2\alpha+\beta\right)\left(\frac{r_0}{r}\right)^{\beta+1}\right],
\end{eqnarray}
%
respectively. 
Equations \eqref{eqrho}--\eqref{eqpt} form a system of three equations for the three unknowns $\rho$, $p_r$, and $p_t$. These three equations are quadratic in their respective unknowns, which implies that this system features a set of at most eight independent solutions, of which some might be complex for given combinations of parameters. 

Due to the complexity of the system of Eqs. \eqref{eqrho}--\eqref{eqpt}, explicit analytical solutions for $\rho$, $p_r$ and $p_t$ are unattainable, even for specific choices of the free parameters $r_0$, $\alpha$, $\beta$, $\gamma$, and $\zeta_0$. Nevertheless, one can obtain analytical solutions for these quantities recursively, as follows. We start by selecting specific values for the free parameters. Then, for this particular choice, starting at $r=r_0$, one can algebraically solve the system for $\rho\left(r_0\right)$, $p_r\left(r_0\right)$, and $p_t\left(r_0\right)$, and obtain the numerical values of these quantities, i.e., sets of values $\left\{\rho_0^i,p_{r0}^i,p_{t0}^i\right\}$, for $i\in\left\{1,...,8\right\}$ corresponding to the eight independent solutions of the system. For each of these solutions, one can increment the radius $r$ in small steps, say $r_{n+1} = r_{n} + \epsilon$ for some small value $\epsilon$, and find the values of $\rho\left(r_{n+1}\right)$, $p_r\left(r_{n+1}\right)$, and $p_t\left(r_{n+1}\right)$. Repeating the process up to a radius $r$ large enough, one is able to extract analytically the behavior of the solutions.

Note that not all of the solutions of the system above are of interest to us. Indeed, we are only interested in those solutions which are relevant from an astrophysical point of view, i.e., solutions whose matter components satisfy the so-called energy conditions. There are four different energy conditions that we are interested in analyzing, namely the null energy condition (NEC), the weak energy condition (WEC), the strong energy condition (SEC), and the dominant energy condition (DEC). For a diagonal energy-momentum tensor $T_{ab}$ in the form of Eq. \eqref{def_matter}, the energy conditions take the forms \cite{Hawking:1973uf}:
\begin{eqnarray}
    &&\text{NEC:} \qquad \rho+p_r \geq 0,\quad \rho+p_t \geq 0,\\
    &&\text{WEC:} \qquad \text{NEC and } \rho \geq 0,\\
    &&\text{SEC:} \qquad \text{NEC and } \rho+p_r+2p_t \geq 0,\\
    &&\text{DEC:} \qquad \rho \geq |p_r|, \quad \rho \geq |p_t|.
\end{eqnarray}
The NEC guarantees that the average energy density as seen by any null observer is positive, the WEC guarantees that the average energy density as seen by any timelike observer is positive, the SEC preserves the attractive behavior of gravity, and the DEC guarantees that the speed of sound is smaller than the speed of light $c$. The DEC is also frequently associated with the stability of the object under study. From the set of eight solutions for the matter quantities contained in the theory, those which violate any of the energy conditions stated above are discarded, and only those which satisfy all of the energy conditions are considered.

\begin{figure*}
  \centering
  \subfigure[\;$\rho$]{\includegraphics[scale=0.8]{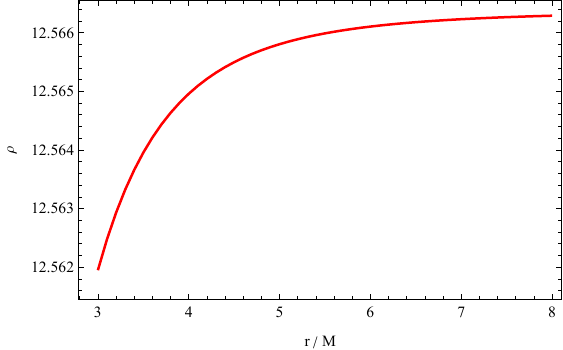}}
  \subfigure[\;$\rho+p_i$, with $i=r,t$]{\includegraphics[scale=0.8]{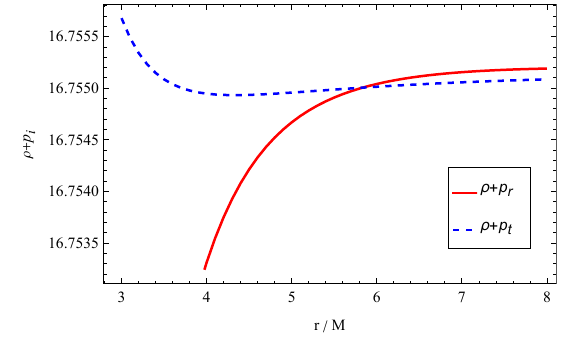}}
  \subfigure[\;$\rho+p_r+2p_t$]{\includegraphics[scale=0.8]{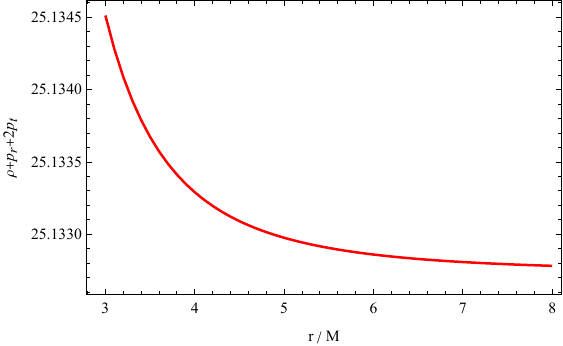}} 
  \subfigure[\;$\rho-|p_i|$, with $i=r,t$]{\includegraphics[scale=0.8]{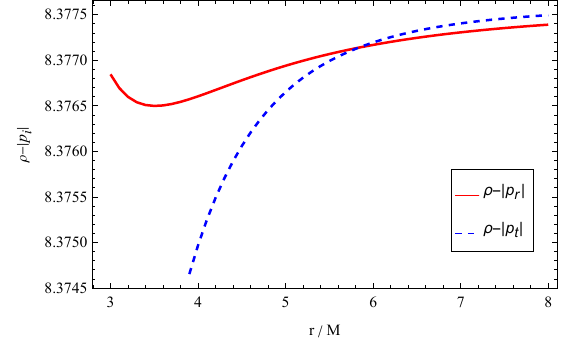}}
  \caption{Matter quantity $\rho$ and the energy condition combinations $\rho+p_r$, $\rho+p_t$, $\rho+p_r+2p_t$, $\rho-|p_r|$ and $\rho-|p_t|$ for the choice of parameters $\alpha=\beta=-\gamma=1$, $r_0=3M$ and $\zeta_0=-\frac{6}{5}$.}
  \label{fig:solution}
\end{figure*}

\begin{figure*}
    \subfigure[\;$\rho$]{\includegraphics[scale=0.58]{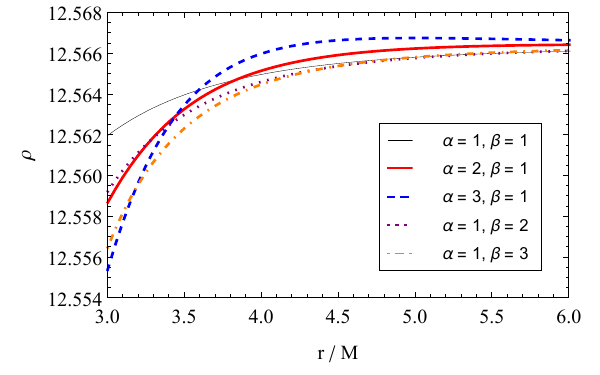}}
  \subfigure[\;$p_r$]{\includegraphics[scale=0.58]{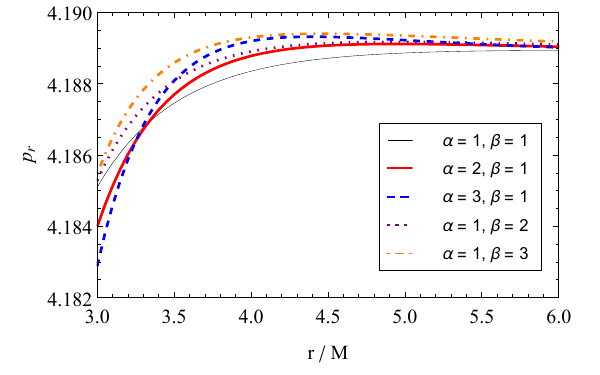}}
  \subfigure[\;$p_t$]{\includegraphics[scale=0.58]{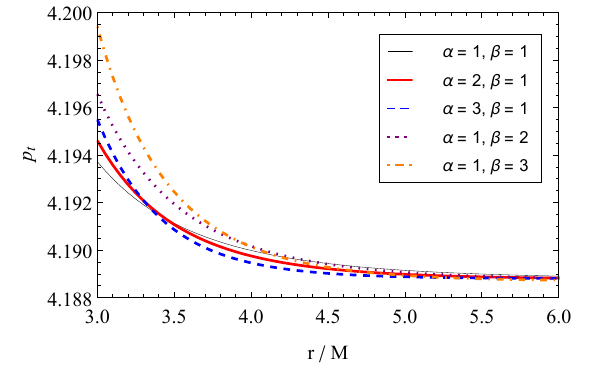}} 
    \caption{Matter quantities $\rho$, $p_r$ and $p_t$ for the choice of parameters $\gamma=1$, $r_0=3M$ and $\zeta_0=-\frac{6}{5}$ and for different combinations of $\alpha$ and $\beta$.}
  \label{fig:solutionab}
\end{figure*}

Similarly to what was previously found in linear $f\left(R,T\right)$ gravity \cite{kull1}, one verifies that in linear 
$f\left(R,\mathcal T\right)$ gravity solutions satisfying all of the energy conditions mentioned above 
exist for negative values of the coupling constant $\gamma$. However, a main difference between 
the two theories is that in the latter the solutions are not asymptotically vacuum, even though they 
are asymptotically flat. As an explicit example, consider the combination of parameters $
\alpha=\beta=-\gamma=1$, $r_0=3M$, and $\zeta_0=-6/5$\footnote{While the choice of $
\zeta_0$ at this point is quite arbitrary, we have chosen this particular value for reasons that we 
clarify in subsequent sections. A wide range of other values of $\zeta_0$, including positive values, 
would provide a qualitatively similar solution.}. The matter quantity $\rho$ and the combinations $
\rho+p_r$, $\rho+p_t$, $\rho+p_r+2p_t$, $\rho-|p_r|$ and $\rho-|p_t|$, for the solution that satisfies 
all of the energy conditions previously mentioned, are plotted in Fig. \ref{fig:solution}. Furthermore, to 
illustrate how the values of $\alpha$ and $\beta$ impact the solutions, we plot the matter quantities 
$\rho$, $p_r$, and $p_t$ for the combination of parameters $\gamma=-1$, $r_0=3M$ and $
\zeta_0=-6/5$ for different combinations of $\alpha$ and $\beta$ in Fig. \ref{fig:solutionab}

An important feature of the solutions described is that, even though the spacetime is asymptotically flat, the matter components are non-vanishing for the whole range of the radial coordinate, i.e., the solutions are not localized. To improve the physical relevance of these solutions, it is thus necessary to consider a matching with an exterior vacuum spacetime at some finite radius, thus resulting in localized wormhole solutions. We address this issue in the following section.

\section{Junction conditions and matching}\label{sec:matching}

\subsection{Junction conditions of linear $f\left(R,\mathcal T\right)$ gravity}

Let the spacetime manifold $\Omega$ be composed of two distinct and complementary regions $\Omega^\pm$, described by metric tensors $g_{ab}^\pm$ written in terms of coordinate systems $x^a_\pm$. We denote the spacetime $\Omega^+$ as the exterior spacetime, and the spacetime $\Omega^-$ as the interior spacetime. The boundary between the two $\Omega^\pm$ is a 3-dimensional hypersurface $\Sigma$ described by a metric $h_{\alpha\beta}$ written in terms of a coordinate system $y^\alpha$, where greek indices exclude the direction orthogonal to $\Sigma$. The projection tensors from the 4-dimensional spacetime $\Omega$ into the hypersurface $\Sigma$ are defined as $e^a_\alpha=\partial x^a/\partial y^\alpha$, and the normal vector on $\Sigma$ is defined as $n_a=\epsilon \partial_a l$, where $l$ is the affine parameter along the geodesics orthogonal to $\Sigma$ and $\epsilon$ is either $1$, $-1$ or $0$ for spacelike, timelike, and null geodesic congruences, respectively. By construction, one has $n^a e_a^\alpha=0$. Following this notation, the induced metric $h_{\alpha\beta}$ and the extrinsic curvature $K_{\alpha\beta}$ of the hypersurface $\Sigma$ can be written as
\begin{equation}\label{defhK}
h_{\alpha\beta}=e^a_\alpha e^b_\beta g_{ab}, \qquad K_{\alpha\beta}=e^a_\alpha e^b_\beta\nabla_a n_b.
\end{equation}

A convenient framework on which to analyze the junction conditions is the distribution formalism. In this formalism, any quantity $X$ and its derivative $\nabla_a X$ can be written in terms of distribution functions as
\begin{equation}\label{distX}
X=X^+\Theta\left(l\right)+X^-\Theta\left(-l\right),
\end{equation}
\begin{equation}\label{distdX}
\nabla_a X=\nabla_a X^+\Theta\left(l\right)+\nabla_a X^-\Theta\left(-l\right)+\epsilon n_a\left[X\right]\delta\left(l\right),
\end{equation}
where $X^\pm$ represent the quantity $X$ in the spacetimes $\Omega^\pm$ respectively, $\Theta\left(l\right)$ is the Heaviside distribution function defined as $\Theta\left(l\right)=0$ for $l<0$, $\Theta\left(l\right)=1$ for $l>0$, and $\Theta\left(l\right)=\frac{1}{2}$ for $l=0$, $\delta\left(l\right)=\partial_l\Theta\left(l\right)$ is the Dirac-delta distribution, and we have introduced the notation $\left[X\right]$ to represent the jump of $X$ across $\Sigma$, i.e.,
\begin{equation}
\left[X\right]=X^+|_\Sigma - X^-|_\Sigma.
\end{equation}
If the quantity $X$ is continuous across $\Sigma$, one has $\left[X\right]=0$. Furthermore, by definition one has $\left[n^a\right]=\left[e^a_\alpha\right]=0$.

To obtain the junction conditions, all of the quantities appearing in the field equations, namely Eq.  \eqref{field}, must be written in the distribution formalism. Consider first the metric $g_{ab}$, which can be written in the distributional formalism as
\begin{equation}\label{eq:def_metric}
g_{ab}=g_{ab}^+\Theta\left(l\right)+g_{ab}^-\Theta\left(-l\right).
\end{equation}
The Christoffel symbols $\Gamma^c_{ab}$ associated with the metric $g_{ab}$ require the computation of the derivatives $\partial_c g_{ab}$. Following Eq.\eqref{distdX}, these derivatives take the form $\partial_c g_{ab}=\partial_c g_{ab}^+\Theta\left(l\right)+\partial_c g_{ab}^-\Theta\left(-l\right)+\epsilon n_c\left[g_{ab}\right]\delta\left(l\right)$. The presence of the term proportional to $\delta\left(l\right)$ is problematic when one tries to define the Riemann tensor $R^a_{\ bcd}$ in the distributional formalism, as the latter depends on products between the Christoffel symbols which consequently depend on factors of the form $\delta^2\left(l\right)$. These factors are singular in the distributional formalism. To avoid these problematic terms, one needs to require that the metric $g_{ab}$ is continuous across $\Sigma$, i.e., $\left[g_{ab}\right]=0$. Since $\left[e^a_\alpha\right]=0$, one can rewrite the previous condition in a coordinate-independent way by projecting both indices into the hypersurface $\Sigma$, from which one obtains
\begin{equation}\label{junction1}
\left[h_{\alpha\beta}\right]=0.
\end{equation}
Equation \eqref{junction1} is known as the first junction condition, and requires the induced metric at $\Sigma$ to be continuous. Following this result, the derivatives of $g_{ab}$ reduce to
\begin{equation}\label{dmetric}
\partial_c g_{ab}=\partial_c g_{ab}^+\Theta\left(l\right)+\partial_c g_{ab}^-\Theta\left(-l\right).
\end{equation}
One is now able to construct the Christoffel symbols in the distribution formalism and, consequently, the Riemann tensor and its contractions, i.e., the Ricci tensor $R_{ab}$ and the Ricci scalar $R$, which are regular. These quantities take the forms
\begin{equation}\label{eq:dist_Rabcd}
R_{abcd}=R^+_{abcd}\Theta\left(l\right)+R^-_{abcd}\Theta\left(-l\right)+\bar R_{abcd}\delta\left(l\right),
\end{equation}
\begin{equation}\label{eq:dist_Rab}
R_{ab}=R^+_{ab}\Theta\left(l\right)+R^-_{ab}\Theta\left(-l\right)+\bar R_{ab}\delta\left(l\right),
\end{equation}
\begin{equation}\label{eq:dist_R}
R=R^+\Theta\left(l\right)+R^-\Theta\left(-l\right)+\bar R\delta\left(l\right),
\end{equation}
where $\bar R_{abcd}$, $\bar R_{ab}$, and $\bar R$ denote the factors proportional to $\delta\left(l\right)$, which are given in terms of geometrical quantities as
\begin{equation}\label{eq:def_barRabcd}
\bar R_{abcd}=4\left[K_{\alpha\beta}\right]e^\alpha_{[a}n_{b]}e^\beta_{[d}n_{c]},
\end{equation}
\begin{equation}\label{eq:def_barRab}
\bar R_{ab}=-\left(\epsilon\left[K_{\alpha\beta}\right]e^\alpha_a e^\beta_b+n_a n_b \left[K\right]\right),
\end{equation}
\begin{equation}\label{eq:def_barR}
\bar R=-2\epsilon\left[K\right],
\end{equation}
where we have introduced the definition of index anti-symmetrization as $X_{[ab]}=\frac{1}{2}\left(X_{ab}-X_{ba}\right)$, and $K=h^{\alpha\beta}K_{\alpha\beta}$ denotes the trace of the extrinsic curvature. 

Let us now consider the matter sector. In general theories of gravity, any terms proportional to $\delta\left(l\right)$ in the gravitational sector of the modified field equations can be associated to the presence of a thin-shell of matter at the hypersurface $\Sigma$. It is thus useful to write the energy-momentum tensor in the distribution formalism as
\begin{equation}\label{eq:dist_tab}
T_{ab}=T_{ab}^+\Theta\left(l\right)+T_{ab}^-\Theta\left(-l\right)+S_{ab}\delta\left(l\right),
\end{equation}
where $S_{ab}=S_{\alpha\beta}e^\alpha_a e^\beta_b$, and $S_{\alpha\beta}$ represents the 3-dimensional energy-momentum tensor of the thin-shell. The scalar $\mathcal T$ in the distributional formalism can thus be obtained via the contraction of $T_{ab}$ with itself, which takes the form
\begin{equation}
\mathcal T=\mathcal T^+\Theta\left(l\right)+\mathcal T^-\Theta\left(-l\right)+\bar{\mathcal T}\delta\left(l\right)+\hat{\mathcal T}\delta^2\left(l\right),
\end{equation}
where the quantities $\bar{\mathcal T}$ and $\hat{\mathcal T}$ are given in terms of matter quantities as
\begin{equation}
\bar{\mathcal T}=\left(T_{ab}^++T_{ab}^-\right)S^{ab},
\end{equation}
\begin{equation}
\hat{\mathcal T}=S_{ab}S^{ab}.
\end{equation}
Similarly to what was previously mentioned to present the regularity of the Riemann tensor, the term proportional to $\delta^2\left(l\right)$ is singular in the distributional formalism and must be eliminated. However, since $\hat{\mathcal T}$ is proportional to a quadratic term in $S_{ab}$, which is always non-negative, the only possible way of eliminating the problematic singular terms in $\mathcal T$ is to force the energy-momentum tensor of the thin-shell to vanish, i.e., 
\begin{equation}\label{junctionS}
    S_{ab}=0.
\end{equation}
Such a matching is called a smooth matching and, while in several other theories of gravity it corresponds to a particular case of the general matching with a thin-shell, in $f\left(R,\mathcal T\right)$ gravity it is the only allowed method of matching two spacetimes that preserves the regularity of the action.

Following the definitions outlined above and under the restriction of Eq. \eqref{junctionS}, the field equations in Eq. \eqref{field} projected into the hypersurface $\Sigma$ with $e^a_\alpha e^b_\beta$ take the form $\left[K_{\alpha\beta}\right]-\left[K\right]h_{\alpha\beta}=0$. Taking the trace of this result with $h^{\alpha\beta}$ implies $\left[K\right]=0$, which upon replacing back into the original equation leads to
\begin{equation}
\left[K_{\alpha\beta}\right]=0,
\end{equation}
i.e., the second junction condition implies that the extrinsic curvature $K_{\alpha\beta}$ must be continuous across $\Sigma$.

Summarizing, the matching between two spacetimes in linear $f\left(R,\mathcal T\right)$ gravity must always be smooth, i.e., in the absence of a thin-shell, and the two junction conditions the spacetimes must satisfy are the same as in GR, that is, the induced metric $h_{\alpha\beta}$ and the extrinsic curvature $K_{\alpha\beta}$ must be continuous across the hypersurface $\Sigma$,
\begin{equation}\label{junction}
\left[h_{\alpha\beta}\right]=0, \qquad \left[K_{\alpha\beta}\right]=0.
\end{equation}

\subsection{Matching with an exterior vacuum}

Let us now make use of the junction conditions derived in the previous section to perform a matching between the interior wormhole spacetime and an exterior spherically symmetric and static vacuum solution. The interior and exterior spacetime metrics to match are 
\begin{equation}\label{metrici}
ds_-^2=-C e^{\zeta_0\left(\frac{r_0}{r}\right)^\alpha}dt^2+\left[1-\left(\frac{r_0}{r}\right)^{\beta+1}\right]^{-1}dr^2+r^2d\Omega^2,
\end{equation}
\begin{equation}\label{metrice}
ds_+^2=-\left(1-\frac{2M}{r}\right)dt^2+\left(1-\frac{2M}{r}\right)^{-1}dr^2+r^2d\Omega^2,
\end{equation}
respectively, where the metric in Eq. \eqref{metrici} corresponds to the metric in Eq. \eqref{def_metric} subjected to the ansatz for the redshift and shape functions given in Eq. \eqref{zbfunctions}, the metric in Eq. \eqref{metrice} corresponds to the Schwarzschild solution with a mass $M$, and the constant $C$ is introduced for later convenience, to guarantee that the time coordinates in both the interior and exterior metrics coincide. 

For a better readability, it is convenient to start the analysis with the second junction condition. Due to the spherical symmetry of the metrics considered, the extrinsic curvatures $K_{\alpha\beta}^\pm$ feature only two independent components, namely $K_{00}$ and $K_{\theta\theta}=K_{\phi\phi}\sin^2\theta$. These components take the forms
\begin{equation}
K_{00}^-=\frac{\alpha\zeta_0}{2r}\left(\frac{r_0}{r}\right)^\alpha\sqrt{1-\left(\frac{r_0}{r}\right)^{\beta+1}},
\end{equation}
\begin{equation}
K_{00}^+=-\frac{M}{r^2}\sqrt{\frac{r}{r-2M}},
\end{equation}
\begin{equation}
K_{\theta\theta}^-=r\sqrt{1-\left(\frac{r_0}{r}\right)^{\beta+1}},
\end{equation}
\begin{equation}
K_{\theta\theta}^+=r\sqrt{1-\frac{2M}{r}}.
\end{equation}

From the second junction condition in Eq. \eqref{junction}, i.e., $\left[K_{\alpha\beta}\right]=0$, one obtains two independent constrains to the matching, namely $\left[K_{00}\right]=0$ and $\left[K_{\theta\theta}\right]=0$. These two constrains take the forms
\begin{equation}\label{cond1}
\frac{\alpha\zeta_0}{2}\left(\frac{r_0}{r}\right)^\alpha\sqrt{1-\left(\frac{r_0}{r}\right)^{\beta+1}}+\frac{M}{r}\sqrt{\frac{r}{r-2M}}=0,
\end{equation}
\begin{equation}\label{cond2}
\sqrt{1-\left(\frac{r_0}{r}\right)^{\beta+1}}=\sqrt{1-\frac{2M}{r}},
\end{equation}
respectively. The second of these conditions, Eq. \eqref{cond2}, can be solved for the radius $r$ and it features a unique real solution for $M>0$ and $r_0>0$. This solution corresponds to the radius $r_\Sigma$ at which the matching must be performed, and it takes the form
\begin{equation}\label{rsigma}
r_\Sigma=\left(2M\right)^{-\frac{1}{\beta}}\left(r_0\right)^{1+\frac{1}{\beta}}.
\end{equation}
Note that the radius $r_\Sigma$ must satisfy the condition $r_\Sigma> 2M$ to avoid the presence of event horizons in the full wormhole spacetime, which according to the result above constrains the radius of the throat to be also in the regime $r_0>2M$, for any $\beta\geq 1$. The solution for $r_\Sigma$ can now be introduced back into the first condition, Eq. \eqref{cond1}, in order to solve it with respect to the value of $\zeta_0$ for which the matching at the radius $r=r_\Sigma$ is possible. Following this procedure, one obtains the solution for $\zeta_0$ as
\begin{equation}\label{zsigma}
\zeta_0=\frac{\left(2M\right)^{\frac{1-\alpha+\beta}{\beta}}\left(r_0\right)^{\frac{\alpha}{\beta}}}{\alpha\left[\left(2M\right)^{1+\frac{1}{\beta}}-\left(r_0\right)^{1+\frac{1}{\beta}}\right]}.
\end{equation}
Note that since $r_0>2M$, which was obtained from the previous constraint, this implies that for any $\alpha\geq 1$ and $\beta\geq 1$, one has $\zeta_0<0$. This is somewhat expected as negative values of $\zeta_0$ preserve the sign of the derivative of $g_{00}$ consistent in both the interior and exterior metrics, a requirement for the matching to be smooth.

Let us now turn to the first junction condition in Eq. \eqref{junction}, i.e., $\left[h_{\alpha\beta}\right]=0$. Since the angular parts of the metrics in Eqs. \eqref{metrici} and \eqref{metrice} coincide, the angular components of the induced metric $h_{\alpha\beta}$ are automatically continuous, and one just needs to analyze the $h_{00}$ components independently. The condition $\left[h_{00}\right]=0$ takes the form
\begin{equation}\label{cond3}
C e^{\zeta_0\left(\frac{r_0}{r}\right)^\alpha}=\left(1-\frac{2M}{r}\right).
\end{equation}
From the analysis of the second junction condition, one has already concluded that the matching must be performed at a radius $r_\Sigma$ given by Eq. \eqref{rsigma}, and that the only value of $\zeta_0$ consistent with this matching is given by Eq. \eqref{zsigma}. Introducing these values of $r_\Sigma$ and $\zeta_0$ into Eq. \eqref{cond3} and solving for the constant $C$ one obtains
\begin{equation}\label{csigma}
C=\left[1-\left(\frac{2M}{r_0}\right)^{1+\frac{1}{\beta}}\right]e^{-\alpha\left[\left(\frac{r_0}{2M}\right)^{1+\frac{1}{\beta}}-1\right]}.
\end{equation}
%
We note that since $r_0>2M$, the constant $C$ is always strictly positive independent of the values of $\alpha\geq1$ and $\beta\geq1$, thus preserving the correct metric signature.

Summarizing, for a given choice of $r_0>2M$, $\alpha\geq 1$, and $\beta\geq 1$, the second junction condition $\left[K_{\alpha\beta}\right]=0$ sets the radius $r_\Sigma$ at which the matching must be performed (see Eq. \eqref{rsigma}) and the corresponding consistent value of $\zeta_0$ (see Eq. \eqref{zsigma}), whereas the first junction condition $\left[h_{\alpha\beta}\right]$ sets the value of the constant $C$ that allows for the complete spacetime metric to be continuous (see Eq. \eqref{csigma}).

\begin{figure*}
    \centering
    \includegraphics[scale=0.8]{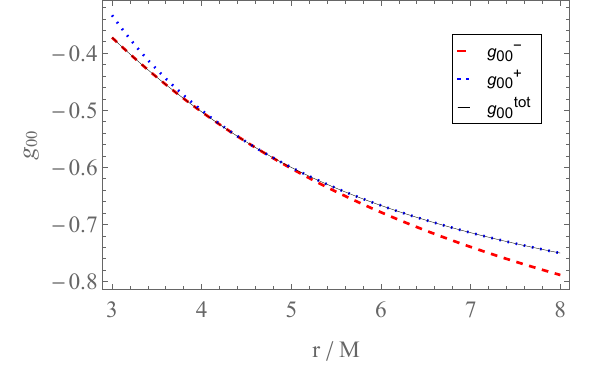}\qquad
    \includegraphics[scale=0.8]{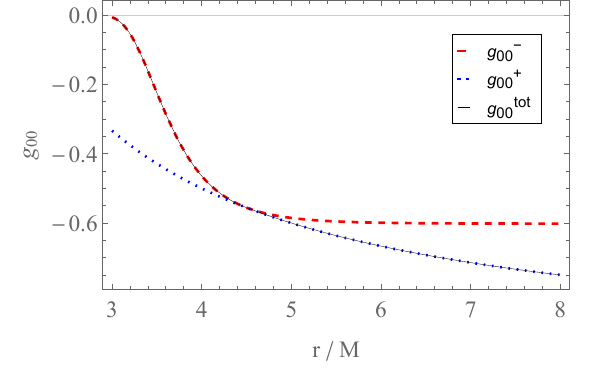}\\
    \caption{Components $g_{00}$ of the interior wormhole spacetime in Eq.\eqref{metrici} (red dashed curve) and the exterior Schwarzschild spacetime in Eq.\eqref{metrice} (blue dotted curve) for $\beta=1$, $r_0=3M$, and $\alpha=1$ (left panel) or $\alpha=10$ (right panel). The thin black line represents the solution $g_{00}^{\text{tot}}$ obtained via the matching between the interior and exterior solutions at $r=r_\Sigma$.}
    \label{fig:matching}
\end{figure*}

Let us provide a couple of explicit examples of application. Consider the particular case $r_0=3M$, $\alpha=1$ and $\beta=1$. For this choice of parameters, Eq. \eqref{rsigma} sets the matching radius at $r_\Sigma=\frac{9}{2}M$, Eq. \eqref{zsigma} sets $\zeta_0=-\frac{6}{5}$, and Eq. \eqref{csigma} sets $C=\frac{5}{9}e^{\frac{4}{5}}$. The $g_{00}$ component of the interior, exterior, and matched metrics are plotted in the left panel of Fig.  \ref{fig:matching}. As an example with a slightly different behavior, consider instead $\alpha=10$ while keeping $\beta=1$ and $r_0=3M$. For this combination, one obtains the same $r_\Sigma=\frac{9}{2}M$, whereas $\zeta_0$ and $C$ take the values $\zeta_0\sim -4.61320$ and $C\sim0.601826$. This solution is plotted in the right panel of Fig. \ref{fig:matching}. We observe that for both solutions the $g_{00}$ component of the metric transitions smoothly from the interior to the exterior metric, thus preserving the continuity of both the induced metric and the extrinsic curvature. 

\begin{figure*}
    \centering
    \includegraphics[scale=0.8]{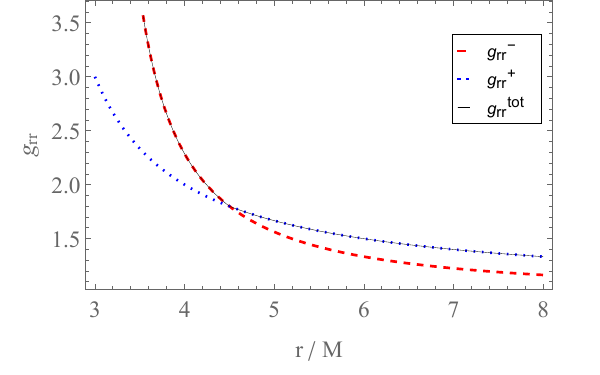}\qquad
    \includegraphics[scale=0.8]{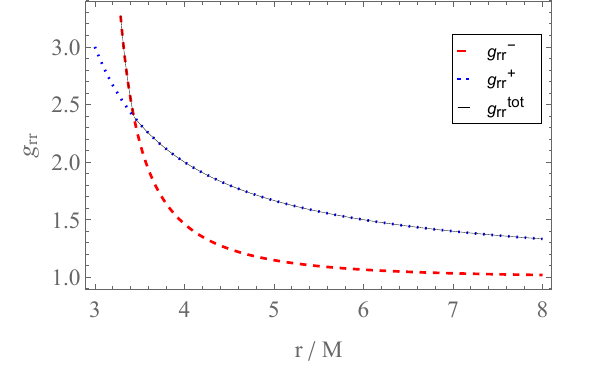}
    \caption{Components $g_{rr}$ of the interior wormhole spacetime in Eq.\eqref{metrici} (red dashed curve) and the exterior Schwarzschild spacetime in Eq.\eqref{metrice} (blue dotted curve) for $\alpha=1$, $r_0=3M$ and $\beta=1$ (left panel) or $\beta=3$ (right panel). The thin black line represents the solution $g_{rr}^{\text{tot}}$ obtained via the matching between the interior and exterior solutions at $r=r_\Sigma$.}
    \label{fig:matchingR}
\end{figure*}

Let us now analyze the radial component of the metric, $g_{rr}$. We take as explicit examples the same case as before for $\alpha=\beta=1$, for which the remaining parameters have already been specified, as well as the combination $\alpha=1$ with $\beta=3$, from which one obtains $r_\Sigma \sim 3.43414 M$, $\zeta_0 \sim -1.59637$, and $C\sim 1.68432$. For both combinations of parameters given, the $g_{rr}$ components of the metric are given in Fig. \eqref{fig:matchingR}. Even though the radial component of the metric is unaffected by the junction conditions, since both the induced metric $h_{\alpha\beta}$ and the extrinsic curvature $K_{\alpha\beta}$ are 3-dimensional tensors on the hypersurface $\Sigma$, one verifies that $g_{rr}$ is continuous, although not differentiable at $r=r_\Sigma$. This continuity of the $g_{rr}$ is expected if one takes into consideration its dependence in the mass function inside a spherical hypersurface of radius $r$, $m\left(r\right)$, i.e., $g_{rr}=\left(1-\frac{2m\left(r\right)}{r}\right)^{-1}$, from which one obtains $m\left(r\right)=\frac{r_0}{2}\left(\frac{r_0}{r}\right)^\beta$ (see Eqs. \eqref{metrici} and  \eqref{zbfunctions}). Indeed, since the matching between the interior and the exterior spacetimes is smooth, i.e., in the absence of a thin-shell, one expects the mass function $m\left(r\right)$ to be continuous at $r_\Sigma$, which implies consequently that the $g_{rr}$ component of the metric is continuous.

\section{Extensions to higher powers of $\mathcal T$}\label{sec:extensions}

In the previous sections, we have analyzed wormhole solutions in a specific form of the $f\left(R,\mathcal T\right)$ theory that is linear in both $R$ and $\mathcal T$. In this form of the action, we have implemented a recursive method to obtain wormhole solutions satisfying the energy conditions, and we have performed a matching with an exterior vacuum spacetime in such a way as to preserve the locality of the solutions obtained. In this section, we argue that as long as the form of the function $f\left(R,\mathcal T\right)$ remains separable and linear in $R$, featuring higher powers of $\mathcal T$, the analysis of the previous sections can be straightforwardly generalized to find wormhole solutions and it is still applicable to these cases, albeit requiring a larger computational time. 

Consider the following extension of the $f\left(R,\mathcal T\right)$ theory used in the previous sections as
\begin{equation}\label{eq:frtnew}
    f\left(R,\mathcal T\right)=R+\gamma \mathcal T + \delta \mathcal T^n.
\end{equation}
For the particular form of $f\left(R,\mathcal T\right)$ given in Eq. \eqref{eq:frtnew}, the field equations and the conservation equation given in Eqs.  \eqref{geo_field} and  \eqref{geo_conservation}, respectively, take the forms
\begin{eqnarray}\label{eq:new_field}
    G_{ab}=8\pi T_{ab}&-&\lambda\left(\Theta_{ab}-\frac{1}{2}g_{ab}\mathcal T\right)-\nonumber \\
    &-&\delta \mathcal{T}^{n-1}\left(n\Theta_{ab}-\frac{1}{2}g_{ab}\mathcal T\right),
\end{eqnarray}
\begin{eqnarray}
    8\pi \nabla_b T^{ab}&=&\gamma\nabla_b\left(\Theta^{ab}-\frac{1}{2}g^{ab}\mathcal T\right)+\nonumber \\
    &+&\delta\nabla_b\left(n\mathcal T^{n-1}\Theta^{ab}-\frac{1}{2}g^{ab}\mathcal T^n\right),
\end{eqnarray}
respectively.

For a metric of the same form as given previously in Eq.  \eqref{def_metric} and a matter distribution of the same form as given in Eq. \eqref{def_matter}, the field equations \eqref{eq:new_field} feature several additional terms in comparison with their linear counterpart, rendering the resultant equations extremely lengthy. As such, we do not write these equations explicitly in this manuscript, but we outline the fundamental differences in what follows.

\subsection{Wormhole solutions}

For a linear form of the function $f\left(R,\mathcal T\right)$ as used in Sec.  \ref{sec:theory}, the field equations \eqref{eqrho}--\eqref{eqpt} are at most quadratic in the matter quantities $\rho$, $p_r$, and $p_t$, which results in a complete set of at most eight possibly complex solutions for these quantities. When one considers higher powers of $\mathcal T$, one consequently obtains higher powers of the matter quantities in the field equations, resulting in a larger set of solutions. Nevertheless, due the fact that the function $f\left(R,\mathcal T\right)$ does not feature any crossed terms in $R$ and $\mathcal T$, the resultant relationship between the matter quantities is still algebraic (in opposition to the differential relation one would find if the crossed terms were present). Consequently, the recursive method outlined in Sec. \ref{sec:worms} to obtain solutions is still applicable. 

\begin{figure*}
  \centering

  \subfigure[\;$\rho$]{\includegraphics[scale=0.8]{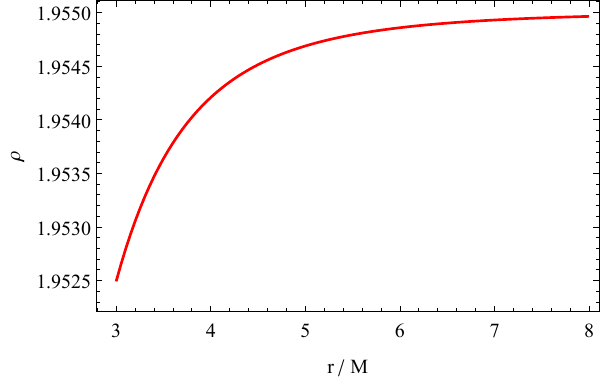}}
  \subfigure[\;$\rho+p_i$, with $i=r,t$]{\includegraphics[scale=0.8]{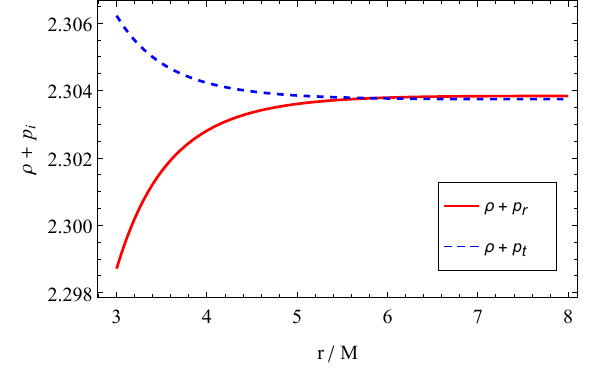}}
  \subfigure[\;$\rho+p_r+2p_t$]{\includegraphics[scale=0.8]{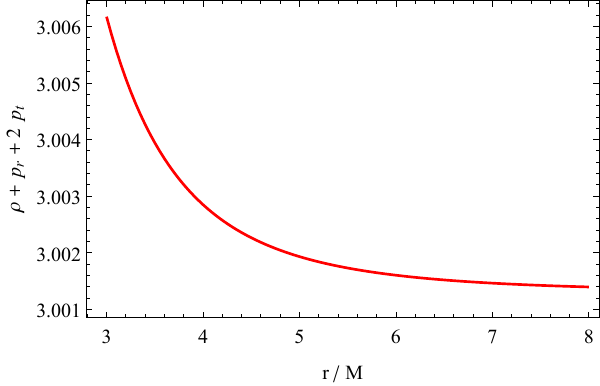}} 
  \subfigure[\;$\rho-|p_i|$, with $i=r,t$]{\includegraphics[scale=0.8]{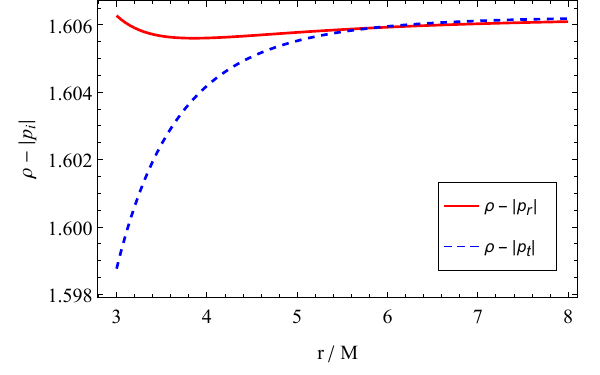}}

  \caption{Matter quantity $\rho$ and the energy condition combinations $\rho+p_r$, $\rho+p_t$, $\rho+p_r+2p_t$, $\rho-|p_r|$ and $\rho-|p_t|$ for the choice of parameters $\alpha=\beta=-\delta=1$, $\gamma=0$, $r_0=3M$ and $\zeta_0=-\frac{6}{5}$.}
  \label{fig:solution2}
\end{figure*}

Similarly to what was found in the linear version of the theory, one verifies that wormhole solutions satisfying all of the energy conditions also exist in theories described by functions $f\left(R,\mathcal T\right)$ featuring higher powers of $\mathcal T$. As an example of application, consider the same wormhole metric as before, described by the parameters $\alpha=\beta=1$, $r_0=3M$, and $\zeta_0=-6/5$, in a particular form of $f\left(R,\mathcal T\right)$ that is linear in $R$ and quadratic in $\mathcal T$, i.e., it is described by the parameters $\gamma=0$, $\delta=-1$, and $n=2$. The matter quantity $\rho$ and the energy condition combinations $\rho+p_r$, $\rho+p_t$, $\rho+p_r+2p_t$, $\rho-|p_r|$, and $\rho-|p_t|$ are given in Fig. \ref{fig:solution2}. Since these solutions are also not localized and require a matching with an external vacuum, we turn to the analysis of the junction conditions in this case in the following section.

\begin{figure*}
  \centering
  \subfigure[\;$\rho$]{\includegraphics[scale=0.58]{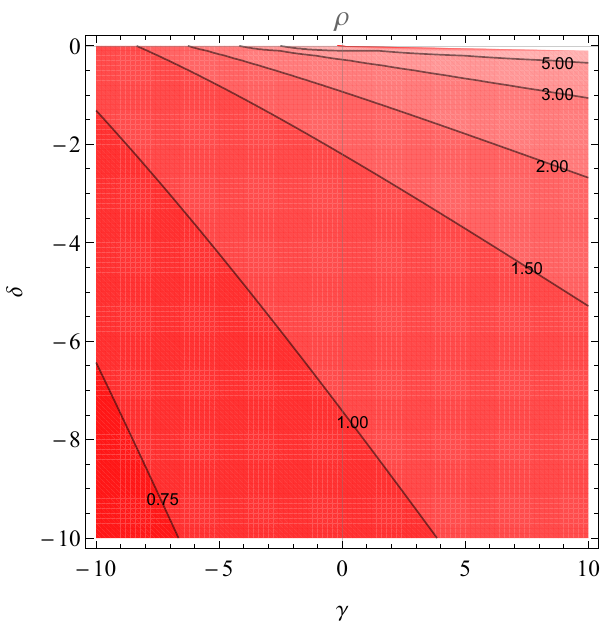}}
  \subfigure[\;$p_r$]{\includegraphics[scale=0.58]{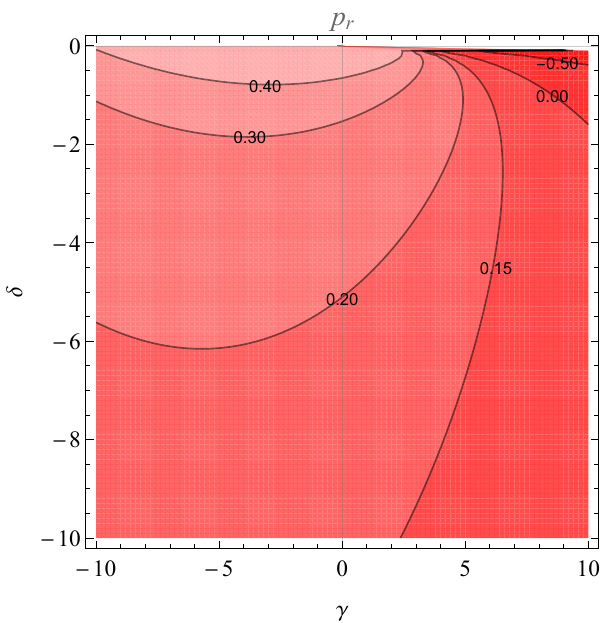}}
    \subfigure[\;$p_t$]{\includegraphics[scale=0.58]{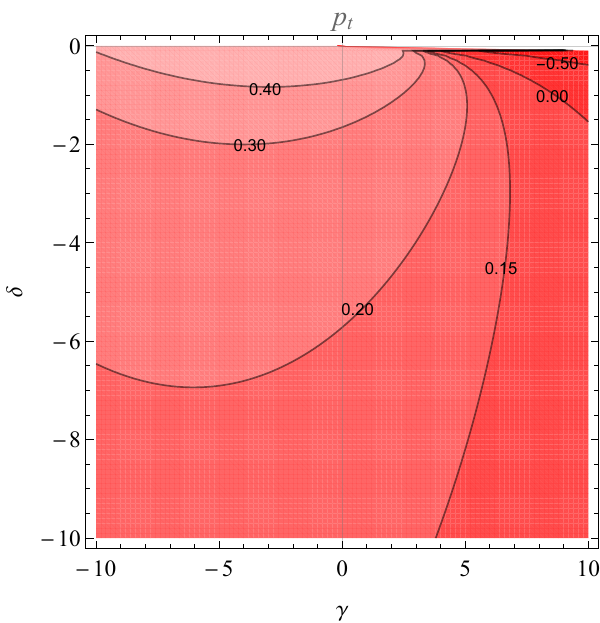}}
  \subfigure[\;$\rho+p_r$]{\includegraphics[scale=0.58]{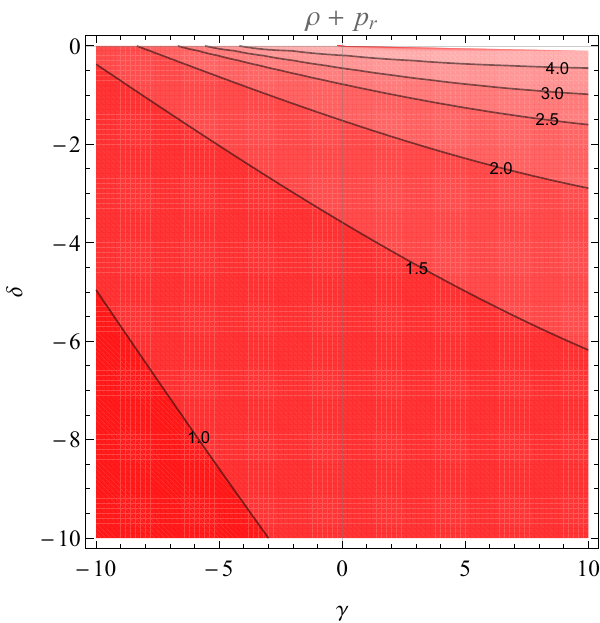}}
    \subfigure[\;$\rho+p_t$]{\includegraphics[scale=0.58]{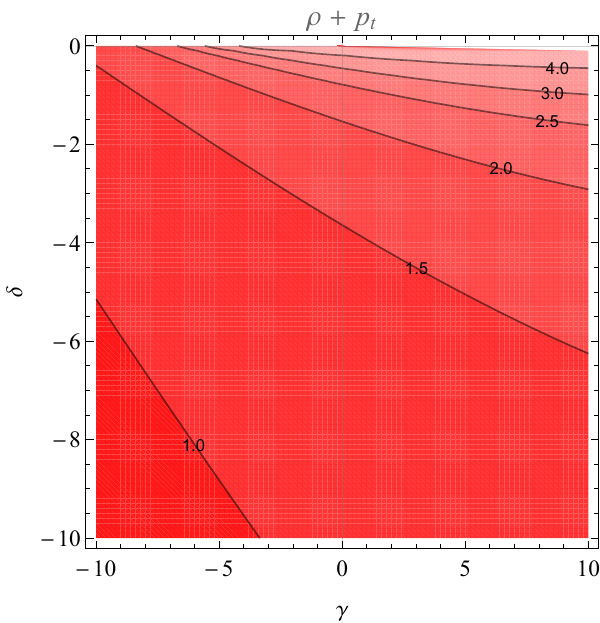}}
      \subfigure[\;$\rho+p_r+2p_t$]{\includegraphics[scale=0.58]{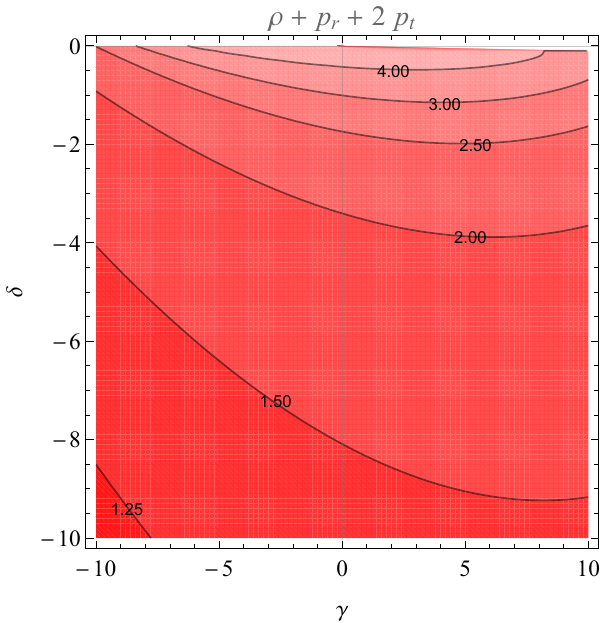}}
  \subfigure[\;$\rho-|p_r|$]{\includegraphics[scale=0.58]{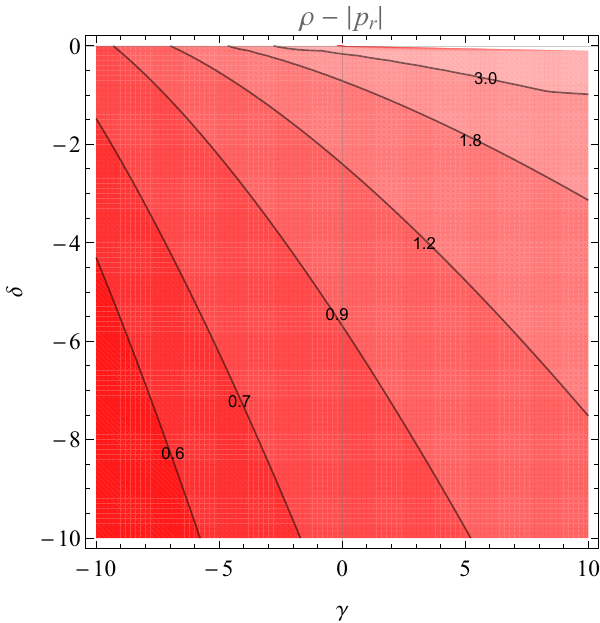}}
    \subfigure[\;$\rho-|p_t|$]{\includegraphics[scale=0.58]{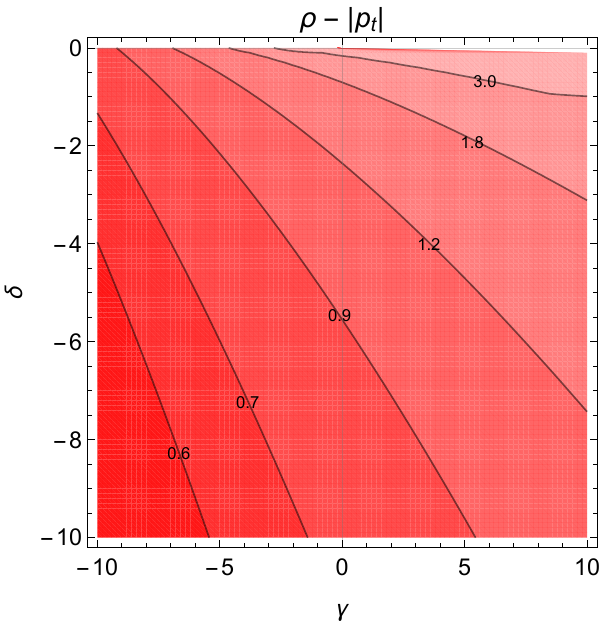}}
  \caption{Values of the matter quantities $\rho$, $p_r$ and $p_t$, as well as the energy condition 
  combinations $\rho+p_r$, $\rho+p_t$, $\rho+p_r+2p_t$, $\rho-|p_r|$, and $\rho-|p_t|$ at the throat 
  $r=_0$ for $\alpha=\beta=1$, $r_0=3M$, $n=2$,  and for different values of $\gamma $ and $
  \delta $.}
  \label{fig:contours}
\end{figure*}

It is also interesting to note that the presence of higher powers of $\mathcal T$ alleviates the restrictions on the lower powers in the form of $f\left(R,\mathcal T\right)$. Indeed, for a linear form of this function, obtained with $\delta=0$, we have verified in Sec. \ref{sec:worms} that solutions satisfying the energy conditions for the whole spacetime could only be obtained by taking negative values of $\gamma$, see e.g. Fig. \ref{fig:solution}. However, when a larger power of $\mathcal T$ is present, one verifies that solutions satisfying the energy conditions can be obtained even for positive values of $\gamma$. In Fig. \ref{fig:contours}, we plot the values of the matter quantities $\rho$, $p_r$ and $p_t$, as well as the energy condition combinations $\rho+p_r$, $\rho+p_t$, $\rho+p_r+2p_t$, $\rho-|p_r|$, and $\rho-|p_t|$, at the throat $r=r_0=3M$ for a function $f\left(R,\mathcal T\right)$ featuring both a linear and a quadratic terms in $\mathcal T$, i.e., $n=2$, for the same wormhole solutions with $\alpha=\beta=1$ and $\zeta_0=-6/5$, with varying values of $\gamma$ and $\delta$. We observe that both the values of $\gamma$ and $\delta$ affect the values of the matter quantities at the throat. However, as long as $\delta$ remains negative, $\gamma$ is allowed to take both positive and negative values, while the solution preserves the validity of all the energy conditions at the throat. Furthermore, all of these solutions that satisfy the energy conditions at the throat also satisfy those conditions for any radius larger than the throat, see e.g. Fig. \ref{fig:solution2} as a particular example.

\subsection{Junction conditions}

In the previous section we have obtained wormhole solutions satisfying the energy conditions for the whole spacetime considering a function $f\left(R,\mathcal T\right)$ that is quadratic in $\mathcal T$. Similarly to the solutions obtained in a linear form of the function, these solutions are not localized and thus require a matching with an external vacuum. Let us analyze what are the consequences of having a higher-order power in $\mathcal T$ in the function $f\left(R,\mathcal T\right)$ to the set of junction conditions obtained in Sec. \ref{sec:matching}.

The modified field equations when a higher-order power of $\mathcal T$ is added to the action are given in Eq. \eqref{eq:new_field}. These equations feature an additional term proportional to $\delta$ in comparison with their linear counterparts, which includes products of the form $\mathcal T^{n-1}\Theta_{ab}$ and powers of $\mathcal T^n$. In Sec.  \ref{sec:matching}, we have proven that in order for $\mathcal T$ to be well-defined in the distributional formalism, it is required that the matching is smooth, i.e., in the absence of a thin-shell, or $S_{ab}=0$. A direct consequence of this restriction is that $\mathcal T$, as well as the auxiliary tensor $\Theta_{ab}$ given in Eq. \eqref{def_theta_2}, are completely regular, i.e., they only feature terms proportional to $\Theta\left(l\right)$ but no terms proportional to $\delta\left(l\right)$. Consequently, the products between $\mathcal T$ and $\Theta_{ab}$, as well as the powers $\mathcal T^n$, also preserve the same regularity. Therefore, no additional junction conditions arise from the addition of a higher-order power-law of $\mathcal T$ in the function $f\left(R,\mathcal T\right)$.

The result of the previous paragraph is based on the same fundamental principle that allows for the relationship between the matter quantities in the modified field equations to remain algebraic, a trait that allowed us to generalize our recursive method to obtain solutions to the case in which higher powers of $\mathcal T$ are present: the fact that crossed terms in $R$ and $\mathcal T$ are absent. Indeed, if one had considered an extension of $f\left(R,\mathcal T\right)$ featuring crossed terms in $R$ and $\mathcal T$, additional differential terms in $\mathcal T$ would appear in the field equations, resulting in additional junction conditions. These extensions will be covered in a separate manuscript, since they are out of the scope of this work.

\section{Conclusions}\label{sec:concl}

In this work, we analyzed traversable wormhole spacetimes in the context of $f\left(R,\mathcal T\right)$ gravity for a linear model on both $R$ and $\mathcal T$. We have proven that a plethora of traversable wormhole solutions satisfying all energy conditions exist in this theory, thus being of a strong physical relevance. Due to the fact that the modified field equations are quadratic in the matter quantities $\rho$, $p_r$, and $p_t$, the theory allows for eight independent (possibly complex) solutions for these quantities, some of which are of limited physical relevance. Thus, we have implemented an iterative recursive algorithm to extract the non-exotic wormhole solutions, and obtain the behavior of the matter quantities.

An interesting feature of the solutions obtained in this work is the fact that, although the spacetime metrics are asymptotically flat, the $f\left(R,\mathcal T\right)$ theory allows for matter distributions that are not asymptotically vacuum, and thus not localized. Nevertheless, the localization of the solutions is possible via the use of the junction conditions of the theory. We have derived such conditions and proven that only a smooth matching is allowed in this theory, as the scalar $\mathcal T$ becomes singular in the presence of a thin-shell. The junction conditions for a smooth matching thus reduce to those of GR, i.e., the continuity of the induced metric and extrinsic curvature at the hypersurface that separates the interior from the exterior spacetime regions. Upon performing the matching mentioned, we have obtained localized wormhole solutions satisfying all of the energy conditions for the whole spacetime, thus being of a particular astrophysical relevance.

The methods introduced in this work can be straightforwardly generalized to more complicated dependencies of the function $f\left(R,\mathcal T\right)$ in $\mathcal T$, provided that crossed terms between $R$ and $\mathcal T$ are absent. Indeed, the fact that the relationship between the matter quantities $\rho$, $p_r$, and $p_t$ remains algebraic in such a case, allows for the extraction of suitable wormhole solutions following the same recursive method. Furthermore, due to the requirement that the matching in this theory is smooth, the absence of these crossed terms also guaranteed that no additional junction conditions arise in the theory, thus allowing one to effectively localize the solutions obtained in precisely the same way as in the linear $\mathcal T$ counterpart. For completeness, we have applied these methods to a quadratic version of the theory and successfully obtained astrophysically relevant and localized wormhole solutions.

The $f\left(R,\mathcal T\right)$ theory of gravity is still a relatively unexplored theory in the topic of wormhole physics due to the complexity of the field equations when more generalized forms of the function are considered. Interesting extensions of this work could cover the analysis of junction conditions for an arbitrary form of the function, particularly including crossed terms of $R$ and $\mathcal T$. Such a form of the function would also require the development of new methods to solve the field equations, as these are promoted to differential relations between the matter fields. An alternative way to address the analysis of more complicated forms of the action is to consider the dynamically equivalent scalar-tensor representation of the theory, in which the arbitrary dependency of the action in $R$ and $\mathcal T$ is exchanged for two scalar fields, with the advantage of reducing the order of the field equations to second-order. We hope to address these issues in upcoming works.

\begin{acknowledgments}
JLR acknowledges the European Regional Development Fund and the programme Mobilitas Pluss for
financial support through Project No.~MOBJD647, and project No.~2021/43/P/ST2/02141 co-funded by the Polish National Science Centre and the European Union Framework Programme for Research and Innovation Horizon 2020 under the Marie Sklodowska-Curie grant agreement No. 94533.
N.G. and F.S.N.L. acknowledge funding from the Funda\c{c}\~{a}o para a Ci\^{e}ncia e a Tecnologia (FCT) research grants UIDB/04434/2020 and UIDP/04434/2020.
F.S.N.L. acknowledges support from the FCT Scientific Employment Stimulus contract with reference CEECINST/00032/2018, and funding from the FCT research grant CERN/FIS-PAR/0037/2019 and PTDC/FIS-AST/0054/2021.
\end{acknowledgments}


\end{document}